\documentclass[conference]{IEEEtran}
\IEEEoverridecommandlockouts
\usepackage{multirow}
\usepackage{float}
\usepackage[draft]{changes}
\usepackage{todonotes}
\definechangesauthor[name={Vahid}, color=red]{VA}
\usepackage{pdfcomment}
\usepackage{amsmath,amssymb,amsfonts}
\usepackage{algorithmic}
\usepackage{graphicx}
\usepackage{textcomp}
\usepackage{xcolor}
\def\BibTeX{{\rm B\kern-.05em{\sc i\kern-.025em b}\kern-.08em
    T\kern-.1667em\lower.7ex\hbox{E}\kern-.125emX}}
\usepackage{url}
\usepackage{hyperref}

\begin{document}

\title{GenXSS: an AI-Driven Framework for Automated Detection of XSS Attacks in WAFs
}

 \author{\IEEEauthorblockN{Vahid Babaey}
 \IEEEauthorblockA{\textit{Department of Electrical and Computer Engineering} \\
 \textit{University of North Carolina at Charlotte}\\
 Charlotte, NC 28223, USA \\
 vbabaey@charlotte.edu}
 \and
 \IEEEauthorblockN{ Arun Ravindran}
 \IEEEauthorblockA{\textit{Department of Electrical and Computer Engineering} \\
 \textit{University of North Carolina at Charlotte}\\
Charlotte, NC 28223, USA \\
 arun.ravindran@charlotte.edu}
 }

\maketitle

\begin{abstract}
The increasing reliance on web services has led to a rise in cybersecurity threats, particularly Cross-Site Scripting (XSS) attacks, which target client-side layers of web applications by injecting malicious scripts. Traditional Web Application Firewalls (WAFs) struggle to detect highly obfuscated and complex attacks, as their rules require manual updates. This paper presents a novel generative AI framework that leverages Large Language Models (LLMs) to enhance XSS mitigation. The framework achieves two primary objectives: (1) generating sophisticated and syntactically validated XSS payloads using in-context learning, and (2) automating defense mechanisms by testing these attacks against a vulnerable application secured by a WAF, classifying bypassing attacks, and generating effective WAF security rules. Experimental results using GPT-4o demonstrate the framework's effectiveness generating 264 XSS payloads, 83\% of which were validated, with 80\% bypassing ModSecurity WAF equipped with an industry standard security rule set developed by the Open Web Application Security Project (OWASP) to protect against web vulnerabilities. Through rule generation, 86\% of previously successful attacks were blocked using only 15 new rules. In comparison, Google Gemini Pro achieved a lower bypass rate of 63\%, highlighting performance differences across LLMs.

\end{abstract}


\section{Introduction}  
\label{sec:introduction}  
The rise of web services has significantly influenced how organizations manage client information. However, as online platforms expand, so do the threats to user data security, with Cross-Site Scripting (XSS) emerging as a prevalent and dangerous attack vector \cite{owasp_xss}. XSS attacks exploit website vulnerabilities to inject malicious scripts, often through links, URLs, or user inputs, which execute within the victim's browser. Web Application Firewalls (WAFs) serve as a primary defense against XSS by detecting and blocking malicious payloads \cite{applebaum2021signature}. However, as web attacks grow more sophisticated, traditional WAFs often struggle to adapt to advanced and obfuscated patterns \cite{mallick2024navigating}. 

Generative AI has emerged as a promising solution, leveraging Large Language Models (LLMs) to generate text, code, and images \cite{abshari2024llm}. LLMs, when combined with techniques like in-context learning and fine-tuning, excel at generating new data from a few relevant examples. \cite{zibaeirad2024comprehensive, abshari2025survey}. This adaptability makes LLMs particularly effective for enhancing WAF defenses by generating simulated attack scenarios based on real-world examples \cite{babaey2025gensqli}. These generated samples can then be analyzed and tested to improve the detection capabilities of WAFs, ultimately contributing to the development of more robust cybersecurity defenses.

In this paper, we propose \textbf{GenXSS}, a generative AI framework for enhancing WAF defenses against XSS attacks. Our framework achieves two primary objectives:  
\begin{enumerate}
    \item \textbf{Generation of Obfuscated XSS Attacks:} Using LLMs and curated in-context learning, GenXSS generates complex XSS payloads validated against real-world vulnerable applications.
    \item \textbf{Automated Defense Mechanisms:} The framework identifies bypassing payloads and employs machine learning and LLMs to generate and validate new WAF rules.
\end{enumerate}  
Experimental results demonstrate the framework's efficacy with 264 new XSS payloads generated, 83\% were syntactically validated using a buggy web application, ensuring they were correctly structured and executable. Among these validated attacks, 80\% bypassed ModSecurity WAF equipped with the latest OWASP Core Rule Set \cite{owasp_modsecurity_crs}, and nearly 100\% bypassed AWS WAF. By classifying the generated payloads using our machine-learning algorithms and generating rules through the GPT-4o LLM rule-generation methodology, we developed and validated security rules that effectively blocked 86\% of all previously successful attacks which bypassed ModSecurity with only 15 new rules. We also employed Gemini Pro LLM to generate new XSS attacks, resulting in a total of 220 generated samples. Of these, 140 (63\%) were validated as syntactically correct and effective, with 104 (74\%) successfully bypassing ModSecurity.

The contributions of this paper include an innovative generative AI framework for XSS mitigation including in-context learning involves incorporating task-specific examples directly within the input prompt to guide the model's behavior and output generation, automated WAF rule optimization, and experimental validation on real-world WAFs. The remainder of the paper is structured as follows: Section \ref{sec:Background} provides a background on XSS attacks and generative AI. Section \ref{sec:Related Work} reviews related literature. Section \ref{sec:genxss_framework} details the framework. Section \ref{sec:evaluation_results} presents results, followed by discussion in Section \ref{sec:discussion}, and conclusions in Section \ref{sec:conclusions}.

\section{Background}
\label{sec:Background}
\subsection{XSS Attacks}
Cross-site Scripting attacks (XSS) can be used by attackers to undermine application security in many ways. XSS vulnerabilities have been used to create social networks worms, steal cookies, spread malware, deface websites, and phish for credentials \cite{owasp_xss}.

Cross-Site Scripting (XSS) attacks come in three main types: Reflected, DOM-based and Stored. Reflected XSS (Non-persistent XSS), the most common type, involves the attacker embedding their payload in a request sent to the web server. The server reflects the payload back in its response, executing it in the victim’s browser. DOM-based XSS is a more advanced, client-side attack where the malicious script exploits vulnerabilities in the web application’s client-side scripts to manipulate the Document Object Model (DOM), executing the payload within the browser. This type of XSS targets elements like URLs or referrers directly in the DOM. Stored XSS (Persistent XSS) occurs when an attacker embeds malicious scripts, such as JavaScript, into a target application. These scripts are typically injected through user input fields, such as comments or posts, and are stored on the server, allowing them to execute whenever other users access the affected content.\cite{owasp_xss}.

\subsection{Web Application Firewalls (WAFs)}
A Web Application Firewall (WAF) is a security mechanism that protects web applications by monitoring and blocking malicious HTTP/S traffic. WAFs typically use two strategies: \textbf{rule-based} and\textbf{ machine-learning-based} approaches. Rule-based methods rely on predefined attack signatures using regular expressions, offering transparency, low false positive rates, and quick deployment, but they are limited to known patterns, require frequent updates, and struggle with scalability. Machine-learning-based methods, on the other hand, detect attacks by learning from data, making them adaptable to novel or obfuscated patterns and capable of detecting anomalies beyond static signatures. However, these methods can have high false positive rates and depend heavily on high-quality labeled data for training and fine-tuning \cite{applebaum2021signature}.

\subsection{Generative AI}
Generative AI is a field of artificial intelligence focused on creating new content, such as text, images, audio, or code, by learning patterns from existing data \cite{zibaeirad2024comprehensive}. Central to this field are Large Language Models (LLMs), such as GPT (Generative Pre-trained Transformer), which use deep-learning architectures to generate coherent and contextually relevant language. These models leverage transformer architectures to capture complex relationships within data and utilize multi-layered neural networks to process vast datasets by extracting increasingly abstract features. With billions or trillions of parameters, LLMs are highly complex and demand substantial computational resources for training.

\section{Related Work}
\label{sec:Related_Work}

While LLMs have shown promise in cybersecurity applications, there are currently limited documented real-world case studies explicitly detailing their deployment in active cybersecurity environments. Most implementations remain in research and experimental phases, focusing on theoretical frameworks or controlled test environments. In this section, we review related work focusing on XSS attack generation, rule adaptation methodologies, and approaches for enhancing WAF robustness.

Wu et al. \cite{wu2024wafbooster} proposed an RNN-based generator for creating malicious payloads, including XSS and SQL injection, validated by a payload corrector. These payloads were tested against a shadow model mimicking WAF behavior and then real-world WAFs. Successfully bypassing payloads were used to enhance WAF detection by updating rules with new signatures.
Yao et al. \cite{yao2023automatic} developed a method for generating XSS attack vectors using an improved Dueling Deep Q-Network (DDQN) algorithm modeled as a Markov Decision Process. Priority experience replay and a reward function based on edit distance guided effective mutations. Generated vectors were validated through semantic analysis and tested against WAFs in both proxy and direct connection modes.
Garn et al. \cite{garn2021combinatorially} employed combinatorial testing with an attack grammar to create diverse XSS payloads, defined by attributes, payloads, and tags. The payloads were tested against multiple WAFs in controlled environments, and their effectiveness was compared to state-of-the-art static attack lists.
Alaoui et al. \cite{alaoui2023generative} used a Generative Adversarial Network (GAN) to create adversarial XSS attacks by modifying existing dataset samples to evade detection by an LSTM-based XSS detection model. The crafted attacks significantly reduced the detection model’s performance, highlighting their effectiveness.
Khan \cite{khan2024ll} introduced a generative AI model combining auto-regressive and transformer techniques to generate XSS payloads by analyzing backend and frontend code. The generated payloads were tested for effectiveness using the OWASP Juice Shop application. 

Our approach differs from traditional methods, where security experts manually update WAF rules in response to observed threats. Furthermore, the works described above present challenges for anomaly detection models that use neural networks. These challenges arise due to the substantial training data required and the complexity of neural network design and implementation. Additionally, we observe that existing XSS attack generation methods in the literature treat attack generation and mitigation as separate processes. In contrast, our framework integrates generative AI to simultaneously automate the generation of XSS attacks and the creation of corresponding WAF rules, offering a unified and adaptive defense mechanism.

\section{GenXSS Framework}
\label{sec:genxss_framework}
In this section, we present the GenXSS framework, a general and flexible solution designed to enhance the performance of WAFs against XSS attacks. The primary advantage of the GenXSS framework lies in its ability to both generate and defend against XSS attacks while offering adaptability to various parameters, including the LLM model, application, and WAF. The framework can be configured to suit different environments based on specific requirements.

\subsection{Architecture}
\begin{figure}[htbp]
\centerline{\includegraphics[width=1\linewidth, height=5cm]{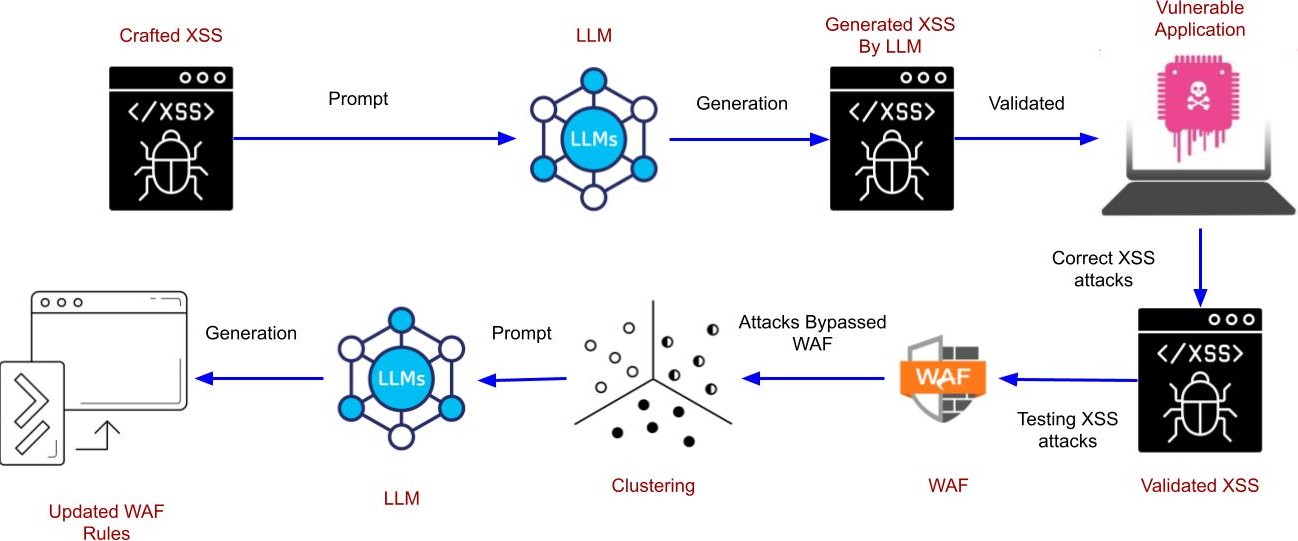}} 
\caption{GenXSS framework architecture.}
\label{fig} 
\end{figure}

In the proposed framework, we utilize in-context learning by providing a set of valid, carefully crafted attacks as examples within the prompt to guide the LLM in generating new XSS samples. However, since LLMs can occasionally produce irrelevant or inaccurate samples, the generated outputs must be validated against a vulnerable application specifically designed to be susceptible to XSS attacks. This validation ensures that only effective and legitimate samples are retained, while irrelevant samples are discarded. Figure \ref{fig} illustrates the architecture of GenXSS.

The next step involves testing the validated samples using a Web Application Firewall (WAF). Samples that successfully bypass the WAF are stored separately and then processed using machine-learning clustering methods. This clustering is a crucial step which group similar attack types for the subsequent generation of security rules to update the WAF. In the second stage of the framework, we use another LLM to generate effective and comprehensive WAF rules tailored to the characteristics of the identified clusters. To maximize rule effectiveness, we apply a Reinforcement Learning with Human Feedback (RLHF) approach. This iterative refinement process ensures the successful mitigation of all XSS attacks.

In the remainder of this section, we describe each of these steps in detail.

\subsubsection{\textbf{XSS Generation}}

In our written prompt, the in-context learning examples were manually crafted and validated against a vulnerable application to ensure they represent real attacks capable of bypassing the WAF. These examples serve as a foundation for the LLM to generate obfuscated XSS attacks. The examples are designed to be complex, clear, and effective to guide the LLM in producing more obfuscated attack samples. The construction of the prompt includes (1) \textbf{Problem description:} A clear explanation of the model's purpose and goal, along with examples provided using few-shot learning. (2) \textbf{Instructions for In-Context Learning:} Systematic obfuscation techniques that guide the LLM to generate new and more obfuscated attacks. (3) \textbf{Tasks:} Precise and detailed tasks that specify exactly what needs to be generated, ensuring the outputs are diverse, comprehensive, and suitable for real-world testing.  

\subsubsection{\textbf{XSS Validation}}
The validation process was conducted using Brute Logic \cite{brutelogic_xss_gym}, a platform specifically designed for testing XSS attacks. XSS examples used for in-context learning were first validated in this application to ensure their accuracy. Brute Logic focuses exclusively on XSS vulnerabilities, providing realistic scenarios such as comment sections, search bars, and URL parameters. It also supports various XSS contexts, including JavaScript event handlers, HTML attributes, and DOM-based injections, making it a versatile tool for comprehensive XSS testing.
Additionally, the XSS examples were validated against a WAF to evaluate their ability to bypass existing defenses. Similarly, LLM-generated XSS attacks were first tested with Brute Logic to ensure syntactic correctness and subsequently with a WAF to verify their ability to exploit vulnerabilities.

\subsubsection{\textbf{XSS Clustering}}
To cluster and analyze the generated XSS attacks, we utilized two machine-learning algorithms: TF-IDF with Hierarchical Agglomerative Clustering (HAC) and SequenceMatcher with Density Based Spatial Clustering of Applications with Noise (DBSCAN). \textbf{TF-IDF + HAC:} The primary strategy of this algorithm is to emphasize unique terms while de-emphasizing common terms. Queries are represented as vectors using TF-IDF, and HAC clusters them based on their similarity \cite{bafna2016document}. For example, -alert\%0d\%0a/**//*test*/(1)- and -alert\%0a/**//*test*/(1)- are grouped due to their shared terms.
\textbf{SequenceMatcher + DBSCAN:} This algorithm calculates character-level similarity scores between queries using SequenceMatcher. DBSCAN then clusters queries with high similarity density \cite{schubert2017dbscan}. For instance, -alert(1)\%0d\%0a//\%20comment- and -alert(1)//comment\%0d\%0a- are grouped due to their structural similarity. The integration of TF-IDF with HAC and SequenceMatcher with DBSCAN provides a robust solution for clustering XSS payloads by leveraging their complementary strengths. TF-IDF + HAC effectively extracts features and captures hierarchical relationships without predefined cluster shapes, while SequenceMatcher + DBSCAN excels in identifying string-level similarities, managing overlapping clusters, and dynamically determining cluster counts. These approaches outperform traditional methods like k-means and regular expression-based clustering.

\subsubsection{\textbf{WAF Security Rule Generation}}
The features of each cluster, along with some of the generated XSS attacks, can be utilized to create a comprehensive and structured prompt for the LLM to generate new security rules for updating WAF security rules. The prompt includes (1) \textbf{System Role:} Assign the LLM the role of a security expert to ensure the generation of efficient security rules specifically designed to block XSS attacks. (2) \textbf{Cluster Characteristics:} Provide a detailed description of each cluster along with a few representative examples. This enables the LLM to analyze the features of each cluster and craft security rules tailored to their specific characteristics. (3) \textbf{Task Definition:} Define the task explicitly to ensure the generated rules are syntactically correct, achieve a high true positive rate, and minimize false positives. The rules are presented as a continuous block for seamless integration, with comments included to explain each rule's purpose and effectiveness.

\subsection{Role of Reinforcement Learning with Human Feedback}
One major challenge in using LLMs is generating accurate outputs, often hindered by unstructured prompts leading to syntax errors and incomplete coverage. Reinforcement Learning with Human Feedback (RLHF) addresses this by iteratively refining prompts based on feedback about issues like redundancy and errors, ensuring improved results over time. Specifically in the GenXSS framework, RLHF enhances rule generation by maximizing true positive rates and minimizing false negatives. and ensuring syntactically correct outputs.

\section{Evaluation and Results}
\label{sec:evaluation_results}
This section evaluates GenXSS based on three aspects: its ability to bypass WAFs, the effectiveness of generated security rules in mitigating attacks, and the adaptability of the methodology to different LLMs.

\subsection{Experimental Setup}
Our experimental setup utilized OpenAI GPT-4o and Google Gemini Pro as the Large Language Models (LLMs) for attack generation. The system operated on Ubuntu 22.04 LTS with Apache v2.4.52 as the web server. For web security, ModSecurity v2.9.5 and AWS WAF were employed as Web Application Firewalls (WAFs), with attack detection rules based on OWASP CRS v4.9.0. The Brute Logic application was used as the vulnerable testing environment, and clustering algorithms are TF-IDF + HAC and DBSCAN + SequenceMatcher for data analysis.
 GPT-4o was configured with a temperature setting of 0.7, and Gemini Pro with a temperature setting of 1. Since our primary goal was to generate diverse yet functional XSS payloads, we relied on these preset values rather than manually optimizing them. While higher temperature typically increases randomness, potentially leading to more invalid payloads, our study did not explicitly measure its effect on validity.

\subsection{Sample XSS Generation}
To illustrate how GenXSS operates, we present two examples of generated XSS attacks: one that is relevant and successfully bypasses the ModSecurity WAF, and another that is irrelevant.

\subsubsection{\textbf{\textbackslash";\textbackslash u0061\textbackslash u006c\textbackslash u0065\textbackslash u0072\textbackslash u0074(1);//}}

This attack which is a relevant one begins with an escape character, a closing double-quote (\texttt{"}), followed by a semicolon (\texttt{;}). This sequence is intended to terminate any existing JavaScript or HTML attribute context prematurely. This is a common technique to escape out of the current context. Next, the query includes Unicode encoding (\textbackslash uXXXX), which represents characters in hexadecimal notation. For instance, \textbackslash u0061\textbackslash u006c\textbackslash u0065\textbackslash u0072\textbackslash u0074 decodes to the JavaScript function \texttt{alert}.

The payload of the attack is the function call \texttt{alert(1)}, which, when executed, displays an alert box with the value \texttt{1}.

Finally, the query includes a comment (\texttt{//}), which indicates the start of a single-line comment in JavaScript. Anything following it on the same line is ignored. This neutralizes any remaining code on the line that could interfere with the attack. The payload is injected using a URL-based attack: https://brutelogic.com.br/gym.php?p16=red\textbackslash \%22;\textbackslash u0061
\textbackslash u006c\textbackslash u0065\textbackslash u0072\textbackslash u0074(1);//

\subsubsection{\textbf{\textbackslash";\textbackslash u0061l\textbackslash x65rt(1);//}}
This payload which is an irrelevant sample fails due to mixed encoding styles \textbackslash u0061l \textbackslash x65rt(1), making the ``alert" function unrecognizable by the JavaScript parser. The inconsistency prevents the script from executing, rendering the payload ineffective.

\subsubsection{\textbf{Bypassing ModSecurity}}  
The first payload bypasses ModSecurity by employing several advanced techniques. First, it utilizes Unicode obfuscation, where Unicode escape sequences (\textbackslash uXXXX) represent characters in the alert function (\textbackslash u0061\textbackslash u006c\textbackslash u0065\textbackslash u0072\textbackslash u0074), effectively evading detection. Second, context breaking is achieved using the \textbackslash "; sequence, which closes an existing string or attribute context and introduces a semicolon (;) to initiate a new statement, disrupting normal parsing mechanisms. Additionally, the payload leverages single-line comments (//) to neutralize any trailing code that might interfere with execution. Finally, it avoids the use of suspicious characters commonly flagged by WAFs, such as \textless, \textgreater, or parentheses (\texttt{()}), further reducing its detectability.

\subsubsection{\textbf{Results}}
Table \ref{tab:XSS_results} summarizes the results of XSS generation using the GPT-4o and Gemini Pro LLM models. Both models were provided with the same prompts along with 4 manually crafted obfuscated XSS samples as in-context examples. The generated samples were validated for correctness using the Brute Logic web application. 
\begin{table}[htbp]
\caption{XSS attack generation results for GPT-4o and Gemini-Pro.}
\label{tab:XSS_results}
\centering
\begin{tabular}{|l|c|c|c|}
\hline
\textbf{LLM Model} & \textbf{Num XSS Attacks} & \textbf{Num Valid} & \textbf{Num Invalid} \\ \hline
GPT-4o            & 264                       & 220                & 44                   \\ \hline
Gemini-Pro        & 220                       & 140                & 80                  \\ \hline
\end{tabular}
\end{table}

\begin{itemize}
    \item GPT-4o: Generated 264 samples, of which 220 were valid XSS payloads, achieving a validity rate of 83\%.
    \item Gemini: Generated 220 samples, of which 140 were valid, resulting in a validity rate of 63\%.
\end{itemize}

Tables \ref{tab:XSS_result_WAF_GPT} and \ref{tab:XSS_result_WAF_Gemini} provide a breakdown of the  performance of ModSecurity WAF against the XSS attacks  generated by GPT-4o and Gemini-Pro respectively. The XSS attacks are listed by type of attack. The generated samples were clustered by type, as each type utilizes a distinct prompt and exhibited unique structures and patterns. The dataset imbalance stems from real-world attack feasibility, as Reflected XSS had four successful manually crafted payloads, while DOM-based XSS had only one, and Stored XSS could not be included due to the lack of a bypassing example.

\renewcommand{\arraystretch}{1.2} 
\setlength{\tabcolsep}{3pt} 
\begin{table}[htbp]
\caption{Validated XSS attacks by type generated by GPT-4o that bypassed ModSecurity.}
\label{tab:XSS_result_WAF_GPT}
\centering
\begin{tabular}{|l|c|c|c|}
\hline
\textbf{Attack Types} & \textbf{Num XSS Attacks} & \textbf{Num Blocked} & \textbf{Num Bypass WAF} \\ \hline
Reflected             & 178                    & 34                 & 144                   \\ \hline
Dom-Based             & 42                     & 12                 & 30                    \\ \hline
\end{tabular}
\end{table}

\begin{table}[htbp]
\caption{Validated XSS attacks by type generated by Gemini Pro that bypassed ModSecurity.}
\label{tab:XSS_result_WAF_Gemini}
\centering
\begin{tabular}{|l|c|c|c|}
\hline
\textbf{Attack Types} & \textbf{ Num XSS Attacks} & \textbf{Num Blocked} & \textbf{ Num Bypass WAF} \\ \hline
Reflected            & 116                       & 22                & 94                   \\ \hline
Dom-Based        & 24                       & 14                & 10                  \\ \hline
\end{tabular}
\end{table}

\begin{itemize}
    \item GPT-4o: Of the 220 validated XSS attacks, 174 (80\%) bypassed ModSecurity.
    \item Gemini Pro: Of the 140 validated XSS attacks, 104 (74\%) bypassed ModSecurity
\end{itemize}

Table \ref{tab:XSS_result_AWS_WAF_GPT} presents the results of GPT-4o generated XSS attacks tested against AWS WAF configured with the Core Rule Set and attached to an Application Load Balancer (ALB). Results indicate that all but one generated attack successfully bypassed the AWS WAF.

\begin{table}[htbp]
\caption{GPT-4o XSS attacks validation that bypassed AWS WAF.}
\label{tab:XSS_result_AWS_WAF_GPT}
\centering
\begin{tabular}{|l|c|c|c|}
\hline
\textbf{Attack Types} & \textbf{ Num XSS Attacks} & \textbf{Num Blocked} & \textbf{ Num Bypass WAF} \\ \hline
Reflected            & 178                       & 1                & 177                   \\ \hline
Dom-Based        & 42                       & 0                & 42                 \\ \hline
\end{tabular}
\end{table}

Table \ref{tab:clustering_results} presents the evaluation results for ModSecurity rules generated using the two clustering techniques described in Section \ref{sec:genxss_framework}. For TF-IDF + HAC, we utilized Ward's linkage metric, which minimizes the variance within clusters, with a distance threshold of 1.8. This approach was chosen to identify a meaningful number of clusters without requiring a predefined cluster count. For DBSCAN, the metric parameter was set using a custom distance matrix, precomputed with a similarity ratio of 1 from SequenceMatcher.
The parameters \texttt{eps = 0.1} and \texttt{min\_samples = 2} were selected after testing, as they produced a minimal yet meaningful set of clusters. To evaluate the cohesion and separation of the clusters, silhouette scores were used, which range from $-1$ to $+1$, with $+1$ indicating strong clustering and $-1$ indicating poor clustering. The silhouette scores for TF-IDF + HAC and SequenceMatcher + DBSCAN were 0.18 and 0.32, respectively, reflecting moderate clustering performance.

\begin{table}[htbp]
\caption{Clustering Results for XSS Attack Types}
\label{tab:clustering_results}
\centering
\scalebox{0.8}{ 
\begin{tabular}{|l|c|l|c|c|}
\hline
\textbf{Attack Types} & \textbf{Num Samples} & \textbf{Clustering Algorithms} & \textbf{Num Rules} & \textbf{Num Blocked} \\ \hline
\multirow{2}{*}{Reflected} & \multirow{2}{*}{144} & TF-IDF + HAC & 9 & 120 \\ \cline{3-5} 
                             &                      & SeqMatcher + DBSCAN          & 5 & 114 \\ \hline
\multirow{2}{*}{Dom-Based}   & \multirow{2}{*}{30} & TF-IDF + HAC                 & 6 & 30 \\ \cline{3-5} 
                             &                      & SeqMatcher + DBSCAN          & 4 & 16 \\ \hline
\end{tabular}
}
\end{table}

The generated clusters were utilized by the LLM for rule generation, with TF-IDF + HAC clustering showing the best performance. Using this approach, GPT-4o generated 15 SecRules that successfully blocked 83\% of the 144 XSS attacks that previously bypassed ModSecurity. Table~\ref{tab:performance_metrics} presents metrics such as False Positive, Precision, Recall, and F1 Score to evaluate the rules' effectiveness in detecting true positives while minimizing errors. For this study, we used a dataset with a 4:1 ratio of normal to attack samples (800 normal and 220 attack samples).

\renewcommand{\arraystretch}{1} 
\setlength{\tabcolsep}{6pt} 
\begin{table}[htbp]
\caption{Performance metrics for generated ModSecurity security rules.}
\label{tab:performance_metrics}
\centering
\begin{tabular}{|l|p{5cm}|}
\hline
\textbf{Metric} & \textbf{Definition/Calculation} \\ \hline
True Positives (TP) & Number of attacks correctly blocked by the WAF: $TP = 150$ \\ \hline
False Negatives (FN) & Number of attacks not blocked by the WAF: $FN = 24$ \\ \hline
True Negatives (TN) & Number of normal samples correctly not blocked by the WAF: $TN = 800$ \\ \hline
False Positives (FP) & Number of normal samples incorrectly blocked by the WAF: $FP = 0$ \\ \hline
Accuracy & $\frac{\text{TN} + \text{TP}}{\text{TN} + \text{TP} + \text{FN} + \text{FP}} = 0.9753$ \\ \hline
Precision & $\frac{\text{TP}}{\text{TP} + \text{FP}} = 1.0$ \\ \hline
Recall & $\frac{\text{TP}}{\text{TP} + \text{FN}} = 0.8621$ \\ \hline
F1-Score & $2 \cdot \frac{\text{Precision} \cdot \text{Recall}}{\text{Precision} + \text{Recall}} = 0.9259$ \\ \hline
\end{tabular}
\end{table}

\section{Discussion} 
\label{sec:discussion}
In this paper, we presented GenXSS, a generative AI-based framework designed to combat highly obfuscated XSS attacks by synthesizing novel attack patterns capable of bypassing Web Application Firewall (WAF) security rules. Unlike traditional manual methods, which require extensive expertise and struggle to address novel patterns, GenXSS integrates machine learning and generative AI to efficiently create scalable defense rules for entire classes of attacks. While generative AI enhances scalability, it also introduces challenges such as computational costs which affect scalability in real-world deployment, token limitations affecting attack diversity, and occasional inaccuracies due to the probabilistic nature of LLMs. Additionally, some publicly available models, such as Anthropic Claude and Meta’s LLaMA, are restricted from generating XSS attack vectors. While these limitations are intended to prevent misuse by attackers, they also hinder defenders from utilizing these models to develop proactive defense mechanisms. The framework’s validation was limited to two types of XSS attacks, a single vulnerable application, and two WAFs ModSecurity and AWS WAF. We also conducted preliminary testing on Cloudflare and Imperva WAFs. These WAFs require domain registration, introducing complexity in controlled experimentation. Also the recall score of 0.86 reflects that while our WAF rule-generation process significantly improved detection, some attacks still bypassed due to limitations in the clustering-based approach. Despite these limitations, GenXSS demonstrates significant potential to enhance cybersecurity defenses through its automated and adaptive rule generation capabilities.

\section{Conclusions and Future Work}
\label{sec:conclusions}

In this paper, we propose a generative AI-based framework
to enhance and secure WAFs against XSS attacks by achieving
two main objectives: generating obfuscated XSS payloads
using LLMs with curated examples validated against real
vulnerable applications, and automating defense mechanisms
through machine learning and LLMs to classify bypassing
attacks and generate new WAF rules. Our experiments demon-
strated the framework’s effectiveness generating 220 new XSS
payloads by GPT-4o, 80\% of which bypassed state-of-the-art
ModSecurity rules. With just 15 new security rules, 83\% of
these unique attacks were successfully blocked.

\subsection{Future Work}
Our framework's results open the door to several promising directions for future research and development. Some limitations, such as recall and dataset imbalance, can be mitigated by manually refining the prompt and iteratively testing it. However, as future work, the plan is to leverage agentic AI to automate this process. By implementing a multi-agent system, each agent will handle a specific task generating, testing, or refining prompts and payloads based on incorrect outputs. These agents will communicate with each other to enhance overall performance \cite{singh2024enhancing}. Furthermore, ethical AI safeguards, including controlled dataset access and responsible disclosure policies, will ensure that generated attack payloads are used exclusively for defensive cybersecurity applications. Another avenue is training and improving machine learning models for anomaly detection by leveraging multi-agent to preprocess training data and design neural network models \cite{et2024dl}.

\bibliographystyle{plain}  
\bibliography{references}

\begin{thebibliography}{10}

\bibitem{abshari2024llm}
Danial Abshari, Chenglong Fu, and Meera Sridhar.
\newblock Llm-assisted physical invariant extraction for cyber-physical systems anomaly detection.
\newblock {\em arXiv preprint arXiv:2411.10918}, 2024.

\bibitem{abshari2025survey}
Danial Abshari and Meera Sridhar.
\newblock A survey of anomaly detection in cyber-physical systems.
\newblock {\em arXiv preprint arXiv:2502.13256}, 2025.

\bibitem{alaoui2023generative}
Rokia~Lamrani Alaoui et~al.
\newblock Generative adversarial network-based approach for automated generation of adversarial attacks against a deep-learning based xss attack detection model.
\newblock {\em International Journal of Advanced Computer Science and Applications}, 14(7), 2023.

\bibitem{applebaum2021signature}
Simon Applebaum, Tarek Gaber, and Ali Ahmed.
\newblock Signature-based and machine-learning-based web application firewalls: a short survey.
\newblock {\em Procedia Computer Science}, 189:359--367, 2021.

\bibitem{babaey2025gensqli}
Vahid Babaey and Arun Ravindran.
\newblock Gensqli: A generative artificial intelligence framework for automatically securing web application firewalls against structured query language injection attacks.
\newblock {\em Future Internet}, 17(1):8, 2025.

\bibitem{bafna2016document}
Prafulla Bafna, Dhanya Pramod, and Anagha Vaidya.
\newblock Document clustering: Tf-idf approach.
\newblock In {\em 2016 International Conference on Electrical, Electronics, and Optimization Techniques (ICEEOT)}, pages 61--66. IEEE, 2016.

\bibitem{brutelogic_xss_gym}
{BruteLogic}.
\newblock {XSS Gym - p04}.
\newblock Available online: \url{https://brutelogic.com.br/gym.php?p04=red} (accessed on 2 December 2024), 2024.

\bibitem{et2024dl}
Maryam Et-Tolba, Charifa Hanin, and Abdelhamid Belmekki.
\newblock Dl-based xss attack detection approach using lstm neural network with word embeddings.
\newblock In {\em 2024 11th International Conference on Wireless Networks and Mobile Communications (WINCOM)}, pages 1--6. IEEE, 2024.

\bibitem{owasp_modsecurity_crs}
OWASP Foundation.
\newblock Owasp modsecurity core rule set project, 2024.
\newblock Accessed: 2024-12-03.

\bibitem{garn2021combinatorially}
Bernhard Garn, Daniel~Sebastian Lang, Manuel Leithner, D~Richard Kuhn, Raghu Kacker, and Dimitris~E Simos.
\newblock Combinatorially xssing web application firewalls.
\newblock In {\em 2021 IEEE International Conference on Software Testing, Verification and Validation Workshops (ICSTW)}, pages 85--94. IEEE, 2021.

\bibitem{khan2024ll}
Sohail Khan.
\newblock Ll-xss: End-to-end generative model-based xss payload creation.
\newblock In {\em 2024 21st Learning and Technology Conference (L\&T)}, pages 121--126. IEEE, 2024.

\bibitem{mallick2024navigating}
Md~Abu~Imran Mallick and Rishab Nath.
\newblock Navigating the cyber security landscape: A comprehensive review of cyber-attacks, emerging trends, and recent developments.
\newblock {\em World Scientific News}, 190(1):1--69, 2024.

\bibitem{owasp_xss}
{OWASP Foundation}.
\newblock {Cross-Site Scripting (XSS)}.
\newblock \url{https://owasp.org/www-community/attacks/xss/}.
\newblock Accessed: 2025-01-07.

\bibitem{schubert2017dbscan}
Erich Schubert, J{\"o}rg Sander, Martin Ester, Hans~Peter Kriegel, and Xiaowei Xu.
\newblock Dbscan revisited, revisited: why and how you should (still) use dbscan.
\newblock {\em ACM Transactions on Database Systems (TODS)}, 42(3):1--21, 2017.

\bibitem{singh2024enhancing}
Aditi Singh, Abul Ehtesham, Saket Kumar, and Tala~Talaei Khoei.
\newblock Enhancing ai systems with agentic workflows patterns in large language model.
\newblock In {\em 2024 IEEE World AI IoT Congress (AIIoT)}, pages 527--532. IEEE, 2024.

\bibitem{wu2024wafbooster}
Cong Wu, Jing Chen, Simeng Zhu, Wenqi Feng, Kun He, Ruiying Du, and Yang Xiang.
\newblock Wafbooster: Automatic boosting of waf security against mutated malicious payloads.
\newblock {\em IEEE Transactions on Dependable and Secure Computing}, 2024.

\bibitem{yao2023automatic}
Yuan Yao, Junjiang He, Tao Li, Yunpeng Wang, Xiaolong Lan, and Yuan Li.
\newblock An automatic xss attack vector generation method based on the improved dueling ddqn algorithm.
\newblock {\em IEEE Transactions on Dependable and Secure Computing}, 2023.

\bibitem{zibaeirad2024comprehensive}
Arastoo Zibaeirad, Farnoosh Koleini, Shengping Bi, Tao Hou, and Tao Wang.
\newblock A comprehensive survey on the security of smart grid: Challenges, mitigations, and future research opportunities.
\newblock {\em arXiv preprint arXiv:2407.07966}, 2024.

\end{thebibliography}
\end{document}